\documentclass[twocolumn,showpacs,preprintnumbers,amsmath,amssymb]{revtex4}

\usepackage{graphicx}
\usepackage{dcolumn}
\usepackage{bm}
\usepackage{amssymb,wasysym}

\begin{document}


\title{Gauge Invariant Photon Mass Induced by Vortex Gauge Interactions}

\author{M. Cristina Diamantini}
\email{cristina.diamantini@pg.infn.it}
\affiliation{%
NiPS Laboratory, INFN and Dipartimento di Fisica, University of Perugia, via A. Pascoli, I-06100 Perugia, Italy
}%

\author{Giuseppe Guarnaccia}
\email{guarnacciagiuseppe@yahoo.it}
\affiliation{%
{Dipartimento di Fisica ``E.~R. Caianiello'', Universit\`a di
Salerno, I-84084 Fisciano (Salerno), Italy}
}%

\author{Carlo A. Trugenberger}
\email{ca.trugenberger@InfoCodex.com}
\affiliation{%
SwissScientific, chemin Diodati 10, CH-1223 Cologny, Switzerland
}%


\date{\today}

\begin{abstract}
We propose a vortex gauge field theory in which the curl of a Dirac fermion current density plays the role of
the pseudovector charge density. In this field-theoretic model, vortex interactions are mediated by a single scalar gauge boson in its antisymmetric tensor formulation. We show that these long range vortex interactions induce a gauge invariant photon mass in the one-loop effective action. The fermion loop generates a coupling between photons and the vortex gauge boson, which acquires thus charge. This coupling represents also an induced, gauge invariant, topological mass for the photons, leading to the Meissner effect. The one-loop effective equations of motion for the charged vortex gauge boson are the London equations. We propose thus vortex gauge interactions as an alternative, topological mechanism for superconductivity in which no spontaneous symmetry breaking is involved. 

\end{abstract}
\maketitle

Topology plays a major role in both field theory and condensed matter systems. It is by now well known that it is not necessary to spontaneously break the U(1) gauge symmetry to generate a photon mass: the famed topological Chern-Simons (CS) term in (2+1) dimensions \cite{jackiw} accomplishes this in a gauge invariant manner, albeit breaking the $P$ and $T$ symmetries. Since the CS term is the infrared-dominant term in the gauge field action it can describe new universality classes of topological matter \cite{wen1} when the dual field strength is used to represent a conserved matter current. 

An analogous mechanism can give photons a gauge invariant mass in (3+1) dimensions, and this time without breaking $P$ and $T$. This is achieved by another topological term, called the $BF$ term and coupling the dual field strength $\tilde F_{\mu \nu}$ to a second-rank antisymmetric pseudotensor $ B_{\mu \nu}$ \cite{bowick}. This is a topological mechanism for superconductivity, in which the field equations for the antisymmetric tensor represent the London equations. 

Due to a second-order gauge symmetry, leaving the action invariant under transformations $B_{\mu \nu}$ $\to $ $B_{\mu \nu} + \partial_{\mu}\lambda_{\nu} - \partial_{\nu}\lambda_{\mu}$, the antisymmetric tensor field encodes actually a single scalar degree of freedom. Classically, this is dual to the phase of a $U(1)$ order parameter and the superconductivity mechanism is dual to the Higgs mechanism in the limit of an infinitely heavy Higgs boson: in the $BF$ mechanism, it is actually the scalar boson that "eats up" the original massless photon to become a massive photon. Quantum mechanically, however, the theories are deeply different, since one can prove \cite{bowick} that no spontaneous symmetry breaking is involved in the $BF$ mechanism, contrary to the Higgs mechanism. The $BF$ superconductivity mechanism is a topological mechanism for superconductivity, genuinely different from the Higgs mechanism. 

It is well known that the CS term in (2+1) dimensions is radiatively induced at one-loop level\cite{redlich} even if it is not present in the original Lagrangian. The generated CS coupling constant $k$, in this case, is proportional to the sign of the fermion mass.

In this paper we show that , also in (3+1) dimensions the $BF$ term is induced at one loop, if the vorticity field of charged fermions is coupled gauge-invariantly to the antisymmetric pseudotensor gauge field $B_{\mu \nu}$. It is known that spin interactions, and particularly spin-orbit coupling, play an important role in the physics of topological insulators \cite{galitski}. It has also been shown that collective excitations, like phonons, can induce long range spin-spin interactions \cite{bennet}. Here we propose analogous vortex gauge interactions as an alternative, topological mechanism for superconductivity. As we will show, in this case, the BF coupling constant is proportional to  $ {m\over 2 \pi}  \ln {\Lambda^2 \over m^2}  = O(m)$, where $m$ is the fermion mass and $\Lambda$ an ultraviolet cutoff. If we compare with the CS case in (2+1) dimensions we see that, in this case, the coupling constant is proportional to the fermion mass. The presence of the ultraviolet cutoff is due to the non-renormalizability of the model in (3+1) dimensions.
Topological mass generation in (3+1)-dimensional QED (without the $B_{\mu\nu}$ tensor field) has been also studied in \cite{dvali} within a generalisation to 4-dimensions of the Schwinger model.

First attempts to extend the gauge principle to vector-like charges \cite{leblanc} have considered as obvious candidates for a tensor gauge theory of spin the standard Dirac tensor and pseudotensor densities $\bar \psi \sigma^{\mu \nu} \psi$ and $\bar \psi \gamma^5  \sigma^{\mu \nu} \psi$, with 
$\sigma^{\mu \nu} \equiv (i/2) [\gamma^{\mu} , \gamma^{\nu}]$ (we shall use units in which $\hbar =1$, $c=1$ throughout the paper, $ [ , ]$ denotes  commutators and  $ \{ , \}$  anticommutators). Both, however are not suitable for a gauge theory, since they are not conserved. 

The spin density of Dirac fermions is given by the expression $\psi^{\dagger} \bf \Sigma \psi$ with $\bf \Sigma = {\rm diag} ({\bf \sigma}, {\bf \sigma})$. Using $\sigma^{ij} = \epsilon^{ijk} \Sigma_k$, this can be embedded in the spin current $S^{\mu , \alpha \beta} = (1/4) \bar \psi \{ \gamma^{\mu}, \sigma^{\alpha \beta} \} \psi$. This, however, is a third-order tensor. In order to obtain a second-order tensor we shall consider its derivative. Moreover, in order to maintain $P$ and $T$ conservation when coupling to the pseudotensor $B_{\mu \nu}$, we shall add a $i\gamma^5$ matrix in the original spin current. We shall thus couple $B_{\mu \nu}$ to the pseudotensor current 
\begin{equation}
J^{\mu \nu} = { i \over 2m} \partial_{\alpha}\left( \bar \psi \gamma^5 \{ \gamma^{\alpha}, \sigma^{\mu \nu} \} \psi\right) =
{- 3\over 2m}  \partial_{\alpha }\left( \bar \psi \gamma^5 \gamma^{ [  \alpha } \gamma^{\mu} \gamma^{\nu ] } \psi \right)\ ,
\label{one}
\end{equation}
where the symbol $[ \dots ]$ in the exponent of [\ref{one}] denotes total antisymmetrization of the indices.
In the second representation of this current, its conservation is explicit. Thus, the coupling $B_{\mu \nu} J^{\mu \nu}$ is invariant under gauge transformations 
\begin{equation}
B_{\mu \nu} \to B_{\mu \nu} + \partial_{\mu} \lambda_{\nu} - \partial_{\nu} \lambda_{\mu} \ .
\label{two}
\end{equation}
It is also invariant under $P$ and $T$ transformations. 

In order to establish what is the pseudovector charge of this gauge theory we must consider the components $J^{0i}$ that couple to $B_{0i}$. By explicit computation we obtain $J^{0i}=- (1/2m) \epsilon^{ijk} \partial_j \ \psi^{\dagger} \alpha^k \psi$. 
The expression $\psi^{\dagger} \alpha^k \psi$ is the velocity field of the Dirac fermion. Thus, in analogy to fluid dynamics, the pseudovector charge density $J^{0i}$, given by the curl of the velocity field, represents the vorticity field of the fermion. The corresponding pseudovector charges $\Phi^i = \int_{\Sigma_i} d^2x J^{0i} n^i$ represent the vortex fluxes through the infinite planes $\Sigma_i$ with unit normals $n^i$. Since $\partial_{\mu} J^{\mu \nu}=0$ these pseudovector charges are conserved.

Using the Gordon decomposition (for the purpose of illustrating the spin dependence of the current, we consider the Gordon decomposition of the  non-interacting case)
\begin{equation}
\psi^{\dagger} \alpha^k \psi = {i\over 2m} \left[ \bar \psi \partial^k \psi - \partial^k \bar \psi \psi \right] 
+{1\over m} \partial_{\mu} \ \bar \psi \sigma^{k \mu} \psi \ ,
\label{three}
\end{equation}
one can separate the pseudovector charges into their orbital and spin contributions. In the non-relativistic limit, in which the lower components of Dirac spinors can be neglected for energies much lower than their mass, this reduces to \cite{greiner} 
\begin{equation}
\left( \psi^{\dagger} \alpha^k \psi \right)_{\rm NR} \to {i\over 2m} \left[ \bar \phi \partial^k \phi - \partial^k \bar \phi \phi \right] + {1\over 2m} \epsilon^{kij} \partial_i  \ \phi^{\dagger} \sigma^j \phi \ ,
\label{four}
\end{equation}
where the spinors $\phi$ on the right-hand side are the two-dimensional Pauli spinor corresponding to the upper components of the four-dimensional Dirac spinors $\psi$. The non-relativistic spin contribution to the pseudovector charge density becomes thus
\begin{equation}
\left( J^{0i}_{\rm spin} \right)_{\rm NR} = {1\over 2m^2} {\nabla}^2 \left( \psi^{\dagger} \sigma^i \psi \right) \ ,
\label{five}
\end{equation}
which is the Laplacian of the intrinsic magnetic moment density of the particle.

We shall henceforth consider a fully relativistic vortex gauge model of massive (mass $m$) charged (charge $e$) Dirac fermions with the following Lagrangian density
\begin{eqnarray}
{\cal L} &&= \bar \psi \gamma^{\mu} \left(i \partial_{\mu} - eA_{\mu} \right) \psi - m \bar \psi \psi
+  i {g\over 9}B_{\mu \nu} J^{\mu\nu} 
\nonumber \\
&&-{1\over 4 } F_{\mu \nu}F^{\mu \nu} + {1\over 12 } H_{\mu \nu \alpha} H^{\mu \nu \alpha} \ ,
\label{six}
\end{eqnarray} 
where $H_{\mu \nu \alpha} \equiv \partial_{[ \mu} B_{\nu \alpha ] } $ is the Kalb-Ramond \cite{kalb} gauge invariant field strength for the antisymmetric tensor and $g$ is the dimensionless vortex coupling constant (the factor $1/9$ has been introduced to simplify later numerical factors). 

It is interesting to note that, using the gamma matrix relation $ \gamma^{ [  \alpha } \gamma^{\mu} \gamma^{\nu ] } = i 6 \epsilon^{\alpha\mu\nu\sigma} \gamma_\sigma$, one can rewrite the vortex interaction $ i {g\over 9}B_{\mu \nu} J^{\mu\nu} $
in (\ref{six}) as:
\begin{equation} i {g\over 9}B_{\mu \nu} J^{\mu\nu}  = - {g \over 2m } F_\mu J^\mu \ ,
\label{fua}
\end{equation}
where $F_\mu = (1/2) \epsilon_{\mu\nu\alpha\beta} \partial^\nu B^{\alpha\beta}$ is the dual Kalb-Ramond field strength and $J^\mu$ is the usual Dirac fermion current. Formally, thus, the dual Kalb-Ramond field strength plays the role of a gauge field in the Lorenz gauge (note however that the corresponding charge $g/2m$ has the dimensions of an inverse mass).

Due to the structure of the Kalb Ramond kinetic term, the mixed components $B_{0i}=-B_{i0}$ play the role of non-dynamical Lagrange multipliers, leaving three dynamical degrees of freedom in the tensor $B_{\mu \nu}$. The gauge invariance eq. (\ref{two}) eliminates, however, two of these, since there are two independent gauge parameters $\lambda_i$ (the other one being eliminated by the equivalence $\lambda_i \equiv \lambda_i + \partial _i \eta $), leaving thus one overall degree of freedom. The long-range vortex gauge interaction is thus mediated by a single massless scalar boson. 

The theory defined by the Lagrangian density ($\ref{six}$) is non-renormalizable due to the mass scale in the denominator of the current density $J^{\mu \nu}$. It is thus to be considered as a low-energy field theory, valid only for energies smaller then the Dirac mass scale $m$. This is, however, exactly the regime in which we are interested, the induced photon mass having the scale $e g m \ll m$ in the perturbative regime $e,g \ll 1$.  In other words we are adopting a hydrodynamic approximation in which we describe long-scale collective phenomena of a relativistic quantum plasma without addressing the high-energy physics on the scale of the individual electrons (positrons) composing it. The vorticity fields $J^{0i}$ that we couple gauge-invariantly to the single scalar embedded in $B_{\mu\nu}$  are typical variables of this regime. As we now show, the induced photon mass is exactly one of the collective phenomena arising at these low energies.

In order to compute the one-loop effective action for the fields $A_\mu$ and  $B_{\mu\nu}$, we rewrite the interaction term between the Kalb-Ramond tensor field and the fermions as:
\begin{equation}L_{\rm int}=\frac{i g}{3m}\bar{\psi}\gamma^5\Gamma^{\rho \mu\nu}\psi\partial_\rho B_{\mu\nu} \ ,
\label{intte}
\end{equation}
where we have performed an integration by parts and used the antisymmetry of $B_{\mu\nu}$; the tensor $\Gamma^{\rho \mu\nu}$ is defined as
\begin{equation}\Gamma^{\rho \mu\nu} = \left( \gamma^\rho\gamma^\mu\gamma^\nu + \gamma^\nu\gamma^\rho\gamma^\mu + \gamma^\mu\gamma^\nu\gamma^\rho \right) \ .
\end{equation}

Integrating over the fermions fields we obtain:
\begin{eqnarray} &&i \Gamma_{\rm eff}(A,B) = \nonumber \\
&&= {\rm Tr} \ln \left( i{\not}\partial -m -e{\not}A+\frac{ i g}{3m}\gamma^5\Gamma^{\rho \mu\nu}\partial_\rho B_{\mu\nu} \right)\ .
\label{onel}
\end{eqnarray}
In (\ref{onel}) we can separate out the trace of the field-independent term $(i{\not}\partial-m)$ and rewrite the resulting logarithm as:
\begin{eqnarray} &i\Gamma_{\rm eff} (A,B) = \nonumber \\
&= {\rm Tr}\ln \left(1+\frac{1}{i{\not}\partial-m} (-e{\not}A +\frac{i g}{3 m}\gamma^5\Gamma^{\rho\mu\nu}\partial_\rho B_{\mu\nu}) \right) \nonumber \\
&= \sum_{n = 1}^{\infty}  {(-1)^{n+1} \over n} {\rm Tr} \left( \frac{1}{i{\not}\partial-m} (-e{\not}A +\frac{ i g}{3 m}\gamma^5\Gamma^{\rho\mu\nu}\partial_\rho B_{\mu\nu})\right)^n \ .
\label{esp}
\end{eqnarray}
The  first term in (\ref{esp}) is the sum of the tadpole diagrams whereas the second term is the relevant quadratic term in the fields $A_\mu$ and  $B_{\mu\nu}$:
\begin{eqnarray} &i \Gamma_{\rm eff}^{(\rm 1-loop)}(A,B) = \nonumber \\
&{-1 \over 2} {\rm Tr} \left[  {1 \over i{\not}\partial-m} \left( - e{\not}A + \frac{i g}{3 m}\gamma^5\Gamma^{\rho \mu\nu }\partial_\rho B_{\mu\nu} \right)\right. \times \nonumber \\
&\times \left. {1 \over i{\not}\partial-m}  \left( - e{\not}A + \frac{i g}{3 m}\gamma^5\Gamma^{\rho \mu\nu }\partial_\rho B_{\mu\nu} \right)\right] \ .
\label{oqua}
\end{eqnarray}

The quadratic term (\ref{oqua}) contains three contributions: the term quadratic  in $A_\mu$, that corresponds to the standard vacuum polarization and gives rise to the renormalization of the electric charge; the quadratic term in  $B_{\mu\nu}$ that gives rise to the corresponding vacuum polarisation for the tensor field and  the interaction term between  $A_\mu$ and $ B_{\mu\nu}$ that, as we now show,  gives rise to the BF term generated at one loop.

In the following we shall use Pauli-Villars regularisation introducing a cutoff  $\Lambda = O(m)$.  Since our theory is an effective theory at long distances, in the calculation we will consistently consider the limit of small momenta $k^2/m^2 <<1$.

The term quadratic in $A_\mu$ is given by:
\begin{eqnarray}
& \Gamma^{(\rm 1-loop)}_{AA} = - \frac{ie^2}{2}Tr(\frac{1}{i{\not}\partial-m}{\not} A\frac{1}{i{\not}\partial-m}{\not} A) \nonumber \\
&= \frac{1}{2}\varint\frac{d^4k}{(2\pi)^4}A_\mu(-k)A_\nu(k)\Pi^{\mu\nu}(k)
\label{AA3}
\end{eqnarray}
where $\Pi^{\mu\nu}$ is the usual vacuum polarization tensor of QED,
\begin{eqnarray}
&\Pi^{\mu\nu}(k)=(g^{\mu\nu}k^2-k^\mu k^\nu)\Pi(k^2)\nonumber \\
& \Pi(k^2)  = \frac{e^2}{12\pi^2}\ln\frac{\Lambda^2}{m^2} \ ,
\label{aa2}
\end{eqnarray}  
for $k^2 << \Lambda^2$.
We obtain thus
\begin{equation}
\Gamma^{\rm 1-loop}_{AA} =  \frac{1}{4}\frac{e^2}{12\pi^2}\ln\frac{\Lambda^2}{m^2} \varint d^4xF_{\mu\nu}F^{\mu\nu} \ .
\label{AA}
\end{equation}

The  computation of the quadratic term in the Kalb-Ramond field is straightforward but more lengthy and tedious since it involves the trace of eight $\gamma$ matrices.
It is given by:
\begin{eqnarray}
&\Gamma^{\rm 1-loop}_{BB} = \frac{i g^2}{9 m^2} Tr(\frac{1}{i{\not}\partial-m}\gamma^5\Gamma^{\mu\nu\alpha}\partial_\alpha B_{\mu\nu}\frac{1}{i{\not}\partial-m}\gamma^5\Gamma^{\rho\sigma\beta}\partial_\beta B_{\rho\sigma})\nonumber \\
&=  \varint\frac{d^4k}{(2\pi)^4}B_{\mu\nu}(k)B_{\rho\sigma}(-k)Q^{\mu\nu\rho\sigma}(k) \ ,
\label{bb1}
\end{eqnarray}
with $Q^{\mu\nu\rho\sigma} (k)$:
\begin{eqnarray}
\label{kernel}
&Q^{\mu\nu\rho\sigma} =   \frac{ig^2}{18m^2}k_\alpha k_\beta\varint\frac{d^4p}{(2\pi)^4} Tr\left(\frac{1}{{\not} p-m}\gamma^5\Gamma^{\mu\nu\alpha} \times \right. \nonumber \\
&\times \left. \frac{1}{({\not}p-{\not}k)-m}\gamma^5\Gamma^{\rho\sigma\beta}\right) \ .
\end{eqnarray}
Using Feynman parametrisation \cite{fey} eq. (\ref{kernel}) can be rewritten as:
\begin{equation}
\label{kernel2}
Q^{\mu\nu\rho\sigma} = -\frac{i g^2}{18m^2}k_\alpha k_\beta\varint_0^1dx\varint\frac{d^4p}{(2\pi)^4}\frac{N^{\alpha \beta \mu\nu\rho\sigma}}{((p-kx)^2-\Delta)^2} \ ,
\end{equation}
with $\Delta=m^2-k^2x(1-x)$, and
\begin{equation}
N^{\alpha \beta \mu\nu\rho\sigma}=Tr(({\not}p+m)\gamma^5\Gamma^{\mu\nu\alpha}({\not}p-{\not}k+m)\gamma^5\Gamma^{\rho\sigma\beta}) \ ,
\label{num}
\end{equation}
that is, the sum of two terms with eight $\gamma$ matrices and one term with six  $\gamma$ matrices. Computing the traces we obtain:
\begin{equation}
Q^{\mu\nu\rho\sigma}(k) = {- g^2 \over 12 \pi^2 } {k^2\over m^2} \ln\left( \frac{\Lambda^2} {m^2}\right) \tilde Q^{\mu\nu\rho\sigma}(k) \ ,
\label{ksk}
\end{equation}
where $\tilde Q^{\mu\nu\rho\sigma}(k)$ is the Kalb-Ramond kernel
\begin{eqnarray}
&\tilde Q^{\mu\nu\rho\sigma}(k) = \left[ k^2 \left( g^{\mu\rho} g^{\nu\sigma} - g^{\mu\sigma} g^{\nu\rho} \right) +  \right. \nonumber \\
&\left. - \left( g^{\mu\rho} k^\nu k^\sigma - g^{\nu\rho} k^\mu k^\sigma \right) -  \left( g^{\nu\sigma} k^\mu k^\rho - g^{\mu\sigma} k^\nu k^\rho \right) \right]  \ .
 \label{kernel4}
\end{eqnarray}
We first notice that, as expected, gauge invariance prevents a mass term for $B_{\mu\nu}$to be  generated at one-loop level, contrary to the result \cite{leblanc}, where the Kalb-Ramond tensor field is coupled to a non -gauge-invariant current. 
This result is crucial, since we are seeking a mechanism for a mass that is topologically generated trough the BF mechanism for the photon in analogy to the gauge invariant mass generation in Chern-Simons theories.
The term eq.(\ref{ksk}), that is generated at the one-loop level is, however, a term of higher order in derivatives with respect to the Kalb-Ramond kernel $\tilde Q^{\mu\nu\rho\sigma}(k)$, due to the non-renormalizability of our model. This term is suppressed in the
 low energy limit in which we are interested, where $m^2 >> k^2$. The coupling constant $g$ is not renormalized at one-loop.

Let us now compute the interaction term between $A_\mu$ and $B_{\mu\nu}$:
\begin{eqnarray}
&\Gamma^{\rm 1-loop}_{BF} = \nonumber \\
&= \frac{  i e g}{3 m}Tr(\frac{1}{i{\not}\partial-m}{\not}A\frac{1}{i{\not}\partial-m}\gamma^5\Gamma^{\mu\nu\rho}\partial_\rho B_{\mu\nu}) \nonumber \\
&= \frac{i e g}{3 m}\varint\frac{d^4k}{(2\pi)^4}A_\sigma(k) B_{\mu\nu}(-k)G^{\mu\nu\sigma}(k)\ ,
\label{ab1}
\end{eqnarray}
with
\begin{equation}
G^{\mu\nu\sigma} (k) =\varint \frac{d^4p}{(2\pi)^4}k_\rho Tr(\frac{1}{{\not}p-m}\gamma^\sigma\frac{1}{({\not}p-{\not}k)-m}\gamma^5 \Gamma^{\mu\nu\rho}) \ .
\label{g1}
\end{equation}
Again, using Feynman parametrization, we have:
\begin{equation}
\label{G}
G^{\mu\nu\sigma}(k) = \varint_0^1dx\varint\frac{d^4p}{(2\pi)^4} \frac{N^{\mu\nu\sigma}}{((p-kx)^2-\Delta)^2} \ ,
\end{equation} 
where:
\begin{eqnarray}
&N^{\mu\nu\sigma}=k_\rho Tr(({\not}p+m)\gamma^\sigma(({\not}p-{\not}k)+m)\gamma^5\Gamma^{\mu\nu\rho})= \nonumber \\
&=-12im^2k_\rho\epsilon^{\mu\nu\rho\sigma}\ .
\label{gg1}
\end{eqnarray}
Using eq. (\ref{G}) and eq. (\ref{gg1}) we obtain the one-loop effective action contribution:
\begin{equation}
\Gamma^{\rm 1-loop}_{BF}  =\frac{ e  g m}{4\pi^2}\ln\frac{\Lambda^2} {m^2}\varint d^4x B_{\mu\nu}(x) \epsilon^{\mu\nu\rho\sigma}\partial_\rho A_\sigma(x) \ .
\label{ab2}
\end{equation}
Eq. (\ref{ab2}) is exactly the BF topological interaction \cite{bowick}, which, as the Chern-Simons term in \cite{jackiw}, is generated at one loop by radiative corrections.

With the standard rescalings $A_\mu \rightarrow e A_\mu$,  $B_{\mu\nu}  \rightarrow g B_{\mu\nu}$ we can write the 1-loop effective  action as:
\begin{equation}
{\cal L}  =  -{1\over 4 e_{\rm ph}^2 } F_{\mu \nu}F^{\mu \nu} + {M\over 2 \pi} B_{\mu\nu} \epsilon^{\mu\nu\rho\sigma}\partial_\rho A_\sigma + {1\over 12 g^2 } H_{\mu \nu \alpha} H^{\mu \nu \alpha}  \ ,
\label{a1l}
\end{equation}
where:
\begin{equation}
e^2_{\rm ph} = e^2 \left( 1 + {e^2 \over 12 \pi^2} \ln {\Lambda^2 \over m^2} \right) \ ,
\label{eph}
\end{equation}
is the renormalised charge (note that this is exactly as in QED since in (3+1) dimensions, contrary to the Maxwell-CS case in (2+1), the Maxwell term is marginal \cite{cri}, and the $B_{\mu\nu}$ field gives no contribution to charge renormalisation at this order) and 
\begin{equation}
M = {m\over 2 \pi}  \ln {\Lambda^2 \over m^2}  = O(m) \ ,
\label{msc}
\end{equation}
is a new mass scale of the order of the electron mass. The action eq.(\ref{a1l}) generalizes to 4-dimensions the Chern-Simons mechanism for  topological mass. The mass is topological  because it does not arises  from spontaneous symmetry breaking, but is due to the presence of the BF term,  independent of the metric. 

The equations of motions are:
\begin{eqnarray}
&\partial_\mu F^{\mu\nu} = - {e_{\rm ph}^2  M \over 6 \pi} \epsilon^{\nu\mu\alpha\beta} H_{\mu\alpha\beta} \ , \nonumber \\
&\partial_\mu H^{\mu\alpha\beta}  = {g^2  M \over 2 \pi} \epsilon^{\alpha\beta\mu\nu} F_{\mu\nu} \ ,
\label{var}
\end{eqnarray}
from which one can derive  \cite{bowick,diam}:
\begin{equation}
\left[ \Box + \left( { M e_{\rm ph}  g \over 2\pi} \right)^2 \right] F_{\mu\nu} = 0 \ .
\label{efm}
\end{equation}
Eq.(\ref{efm}) shows that the field strength $F_{\mu\nu} $ satisfies the Klein-Gordon equation with mass given by ${ M e_{\rm ph}  g \over 2\pi} $. By the same analysis also $H^{\mu \nu \alpha} $ satisfies the same equation of motion.

From eq.(\ref{var}) we also recognize that the dual of the Kalb-Ramond field strength acts as charged current for the photon field:
\begin{equation}
\partial_\mu F^{\mu\nu} =  J^\nu\ ,   J^\nu \equiv  {- e_{\rm ph}^2  M \over  \pi}H^\nu =  {- e_{\rm ph}^2  M \over 6 \pi} \epsilon^{\nu\mu\alpha\beta} H_{\mu\alpha\beta}  \ .
\label{var1}
\end{equation}
Substituing eq.(\ref{var1}) in the second equation  eq.(\ref{var}) we obtain:
\begin{eqnarray} 
&\epsilon^{\mu\nu\alpha\beta} \partial_\alpha J_\beta =  - \left( {e_{\rm ph} g M \over  \pi} \right)^2  \tilde F_{\mu\nu}  \ , \nonumber \\
&\tilde F_{\mu\nu} = {1\over 2} \epsilon^{\mu\nu\alpha\beta} F_{\alpha\beta}
\label{var2}
\end{eqnarray}
which is nothing else than a relativistic version of the London equations for superconductivity.   

In conclusion, we have shown that electron vorticity gauge interactions mediated by a scalar boson in a hydrodynamic effective low-energy field theory generate the BF topological mass term at 1-loop in the effective action. 
Photons acquire a gauge-invariant topological mass, the additional scalar boson acquires charge and the corresponding current satisfies the London equations, all the hallmarks of "gauge-invariant" superconductivity  with no spontaneous symmetry breaking involved.

We thank Prof. H. Hansson for  a critical reading and very helpful comments on the manuscript.


\begin{references}
\bibitem{jackiw}R. Jackiw and S. Templeton, {\it Phys. Rev.} {\bf D23} (1981) 2291; 
S. Deser, R. Jackiw and S. Templeton, {\it Phys. Rev. Lett.} {\bf 48} (1982) 975;  {\it Ann. Phys.} (N.Y.) {\bf 140} (1982) 372.
\bibitem{wen1}For a review see e.g.: X.-G. Wen, "Topological Order: From Long-Range Entangled Quantum Matter to an Unification of Light and Electrons", arXiv:1210.1281. 
\bibitem{bowick}T. J. Allen, M. Bowick and A. Lahiri, {\it Mod. Phys. Lett.} {\bf A6} (1991) 559; M. Bergeron, G. Semenoff and R. J. Szabo, {\it Nucl. Phys.} {\bf B437} (1995) 695. 
\bibitem {redlich}A. N. Redlich, {\it Phys. Rev. Lett. } {\bf 52} (1984) 18; {\it Phys. Rev.} {\bf D29} (1984) 236.
\bibitem{galitski}See e.g.: V. Galitski and B. Spielman, {\it Nature} {\bf 494} (2013) 49. 
\bibitem{bennet}S.D. Bennet, N. Y. Yao, J. Otterbach, P. Zoller, P. Rabl and M. D. Lukin, {\it Phys. Rev. Lett.} {\bf 110} (2013) 156402.
\bibitem{dvali}G. Dvali, R. Jackiw and S-Y. Pi,  {\it Phys. Rev. Lett.} {\bf 96} (2006) 081602.
\bibitem{leblanc}M. Leblanc, R. MacKenzie, P. K. Panigrahi and R. Ray, {\it Int. J. Mod. Phys.} {\bf A09} (1994) 4717. 
\bibitem{greiner}See e.g.: W. Greiner, "Quantum Mechanics, 4th ed.", Springer Verlag, Berlin (2001); M. Nowakowski, {\it Am. J. Phys.} {\bf 68} (2000) 259. 
\bibitem{kalb}M. Kalb and D. Ramond, {\it Phys. Rev. } {\bf D9} (1974) 2273.
\bibitem{fey}See e.g.: S. Weinberg, { \it "The Quantum Theory of Fields"} Vol. 1, Cambridge University Press.
\bibitem{diam}M.C. Diamantini, P. Sodano and C.A. Trugenberger, {\it  Eur.Phys.J.} {\bf B53} (2006) 19.
\bibitem{cri} M.C. Diamantini, P. Sodano and C.A. Trugenberger,  {\it New Journal of Physics} {\bf 14} (2012) 063013.
\end{references}
\end{document}